\documentclass [11pt,a4paper] {article}

\usepackage[cp1252]{inputenc}

\usepackage{amssymb}
\usepackage{amsmath}
\usepackage{amsfonts,amssymb}

\usepackage[dvips]{graphicx}

\usepackage{bbm}

\DeclareMathAlphabet{\mathpzc}{OT1}{pzc}{m}{it}

\begin{document}

\title{Quantum probabilities and the Born rule in the intuitionistic interpretation of quantum mechanics}

\author{Arkady Bolotin\footnote{$Email: arkadyv@bgu.ac.il$\vspace{5pt}} \\ \textit{Ben-Gurion University of the Negev, Beersheba (Israel)}}

\maketitle

\begin{abstract}\noindent This paper presents a novel explanation of the cause of quantum probabilities and the Born rule based on the intuitionistic interpretation of quantum mechanics where propositions obey constructive (intuitionistic) logic. The use of constructive logic makes it possible (through a replacement of the concept of truth with the concept of constructive probability) to abandon the law of excluded middle in the intuitionistic interpretation.\\

\end{abstract}

\section{Introduction}\label{Introduction}

\noindent Whence comes indeterminism in physics? Really, assuming that the Schr\"odinger dynamics is universally valid, i.e., that each process in physics is governed by the (linear differential) Schr\"odinger equation and so is continuous, causal, and reversible, how is that there are processes involved in measurement that are indeterministic, i.e., whose outcomes are probabilities? It is no exaggeration to say that the problem this question relates to is the most difficult in the foundation of quantum mechanics.\\

\noindent Under the Copenhagen interpretation, quantum probabilities enter physics through the Born rule, which cannot be investigated as it is a postulate of the theory (the same one as the Schr\"odinger equation is). However, the problem with such an \textit{ab initio} postulation is not that it does not offer any deeper understanding of how probabilities come into existence. The real problem here is that there is nothing inherent in the Copenhagen interpretation that can either reject the mutual application of the Schr\"odinger equation and the Born rule to a system of interest or select one postulate over another \cite{Tammaro}.\\

\noindent Hence, attempts to understand the cause of indeterminism in quantum physics are being made beyond the Copenhagen interpretation.\\

\noindent For example, in the relative state (also referred to as many-worlds) interpretation (henceforward RS/MWI for short), the behavior of a rational agent in quantum-mechanical situations is thought to provide a natural account of the probability concept \cite{Deutsch,Barnum,Wallace,Gill,Assis}. As it is argued in the quantum-mechanical version of decision theory, being strongly constrained in their behavior the rational agents would quantify their subjective uncertainty in the face of the wave function ``splitting'' (or ``branching'') by the use of probability given by the Born rule.\\

\noindent But then the fact that the RS/MWI contains neither element nor quality nor attribute related to a notion of \textit{belief} makes a doubt if there be a way to derive probabilities – i.e., \textit{degrees of beliefs} – from such an interpretation.$\,$\footnote{\label{f1}The concept of an agent's degree of confidence, \textit{a graded belief}, is one of the main concepts of probability. See papers \cite{Eriksson,Hajek09} that analyze in depth this and other concepts of probability.\vspace{5pt}} What is more, even if this notion had been added to the RS/MWI, it still would be flatly inconsistent with the universally applicable (deterministic) Schr\"odinger dynamics and the splitting picture \cite{Hemmo}. Such an argumentation may explain why no decisive conclusion on the possibility of emergent probabilities and randomness in the RS/MWI has been reached yet.\\

\noindent Meanwhile, the decoherence program tries to find the solution to the problem of the origin of quantum probability by including the environment but without relying on the key elements of decoherence that presume the Born rule and would thus render the contention circular \cite{Zeh96,Zeh03}. For example, as it has been proposed in the decoherence program, the Born rule is originated from the environment-assisted invariance (in other words, from a symmetry of composite quantum states) \cite{Zurek,Schlosshauer,Mohrhoff}.\\

\noindent But then again, it hard to see how probabilities can emerge within the framework, in which only unitary evolutions (without collapse mechanism) are allowed and thus there do not exist measurement outcomes, that is, \textit{the bearers of probabilities}.$\,$\footnote{\label{f2}According to the propensity, subjective, and logical interpretations of probability, measurement outcomes (or events) are the bearers of probabilities; see, for example \cite{Hajek12,Easwaran}.\vspace{5pt}}\\

\noindent So, the intention of this paper is to give explanation of the cause of quantum probabilities and the Born rule using a completely different approach. The key idea is to employ the intuitionistic interpretation of quantum mechanics \cite{Bolotin} in which propositions concerning quantum-mechanical situations obey constructive (intuitionistic) logic. As the objects of belief that bear logical values (`true' $\top$ and `false' $\bot$) \cite{McGrath}, propositions let one pave the way for the entrance of a probabilistic concept into quantum theory. Also, the use of intuitionistic logic makes it possible (through a replacement of the concept of truth with the concept of constructive provability) for the assumption of the universal validity of the Schr\"odinger dynamics to \textit{become weaker}. Such a deregulation allows one to drop \textit{the law of excluded middle} in the intuitionistic interpretation so that the last-named does not ``fall victim to Schr\"odinger's cat and the like'' \cite{Caspers}.\\

\noindent The rest of the paper is structured as follows: The essence of the points maintained in the paper is put in the next Section \ref{Probabilities}, while the Section \ref{Conclusion} concludes the arguments of this paper.\\

\section{How irreducible probabilities appear in the intuitionistic interpretation}\label{Probabilities}

\noindent Assume that a (typical microscopic) system $S$, whose pure quantum states are determined by the eigenstates $|{a_1}\rangle$ and $|{a_2}\rangle$ of the observable $A$ (that takes on only two possible values $a_1$ and $a_2$), interacts for the duration of the measurement $t$ with a (typical macroscopic) apparatus $M$ designed to measure (observe) $A$.\\

\noindent Suppose that the state of the system $S$ is a superposition $|{\psi}\rangle = {c_1}|{a_1}\rangle+{c_2}|{a_2}\rangle$ where coefficients $c_1$ and $c_2$ meet with the normalization condition ${c_1}^2+{c_2}^2=1$ (whose reason will be explained later). In this case, the sample space of the measurement  -- i.e., the set of all possible outcomes of the measurement of the observable $A$ -- can be defined as ${\Omega} \equiv \{a_1,a_2\}$. Let us evaluate the following proposition:$\,$\footnote{\label{f3}In this paper, propositions and statements are used interchangeably and denoted (if not stated otherwise) as enclosed in parentheses expressions $(\cdot)$ which are capable of being true or false.\vspace{5pt}}\smallskip

\begin{equation} \label{1} 
        \left(\{a_1,a_2\}\right)
        \equiv
        \big(
           \left(\{a_1\}\right) \oplus \left(\{a_2\}\right)
        \big)
        \equiv
        \Big(
           \big(
              \left(\{a_1\}\right) \vee \left(\{a_2\}\right)
           \big)
           \wedge
           \big(
              \neg \left(\{a_1\}\right) \vee \neg \left(\{a_2\}\right)
           \big)
        \Big)   
   \;\;\;\;  ,
\end{equation}
\smallskip

\noindent where $\oplus$ stands for the logical operation of exclusive disjunction (corresponding to the construct ``either ... or'') that outputs `true' only when its inputs -- i.e., the propositions of elementary events $\{a_1\}$ and $\{a_2\}$ (subsets of $\{a_1,a_2\}$) -- differ, namely, $(\{a_1\}) \neq (\{a_2\})$. The proposition $(\{a_1,a_2\})$ is logically equivalent to the assertion that there is a final (i.e., at the moment $t$ in the last part of the measurement) state $|{\Psi}_t\rangle$ of the composite system $S+M$ in which the observable $A$ has a definite value, in other words, there is an event which contains only a single outcome -- \textit{either} $\{a_1\}$ \textit{or} $\{a_2\}$ – in the sample space $\{a_1,a_2\}$ of the measurement.\\

\noindent Consider the first part of the compound expression (\ref{1}), that is, the proposition $\left( (\{a_1\}) \vee (\{a_2\}) \right)$. In view of the fact that this proposition contains the logical constant $\vee$, constructive logic \cite{Artemov} (adopted in the intuitionistic interpretation of quantum mechanics presented here) requires that if this proposition is asserted to be true, then \textit{witness} must be given, which chooses $(\{a_1\})$ or $(\{a_2\})$ and provides (an explicit, constructive) proof for it.$\,$\footnote{\label{f4}Another logical constant whose interpretation requires constructive proof is $\exists$.\vspace{5pt}} What counts as ``witness'' is open to interpretation though. In the said case, witness can be understood as an outcome of either \textit{an actual measurement of $A$} or a decision problem that determines the logical value of the proposition $\left( (\{a_1\}) \vee (\{a_2\}) \right)$ \textit{before the actual measurement of $A$}.\\

\noindent This decision problem can be defined as the following propositional formula:\smallskip

\begin{equation} \label{2} 
        \big(
           \left(\{a_1\}\right) \vee \left(\{a_2\}\right)
        \big)
        \equiv
        \left(
           |{\Psi}_t\rangle
           =
          |{a_1}\rangle  |{M_1}\rangle
          \vee
          |{\Psi}_t\rangle
           =
          |{a_2}\rangle  |{M_2}\rangle
        \,\Big|\,
          P(|{\Psi}_t\rangle)
        \right)
   \;\;\;\;  ,
\end{equation}
\smallskip

\noindent where $|{M_1}\rangle$ and $|{M_2}\rangle$ represent mutually orthogonal quantum states of $M$, both orthogonal to $|{M_0}\rangle$ (which is the initial -- i.e., at the moment $t=0$ preceding the measurement -- quantum state of $M$, so-called \textit{ready state}), corresponding to different macroscopic configurations of $M$ similar to different positions of a pointer along a scale, while $P(|{\Psi}_t\rangle)$ represents a statement -- i.e., a predicate -- that may be true or false depending on the particular final state $|{\Psi}_t\rangle$ of the composite system $S+M$.\\

\noindent By contrast, when the state of the system $S$ is just one of the eigenstates $|{a_n}\rangle$ of the observable $A$, that is, $|{\psi}\rangle = |{a_n}\rangle$ (where $n \in \{1,2\}$), the sample space of the measurement correspondingly consists of a single element, i.e., ${\Omega} \equiv \{a_n\}$. In the given case, a witness for the proposition of the elementary event $(\{a_n\})$ (asserting that in the final state $|{\Psi}_t\rangle =|{a_n}\rangle |{M_n}\rangle$ the observable $A$ has the definite value $a_n$) \textit{is required to do nothing but only exist if $(\{a_n\})$ is true}. Therefore, the proposition $(\{a_n\})$ can be treated in accordance with the laws of classical logic \cite{Taranovsky}. In particular, the following propositional expression\smallskip

\begin{equation} \label{3} 
        (\{a_n\})
        \equiv
        \left(
          |{a_n}\rangle |{M_n}\rangle
        \right)
        =
        \top
   \;\;\;\;   
\end{equation}
\smallskip

\noindent (where symbol = positioned after the parentheses defines the notion of evaluation) can be taken as a postulate assuming \textit{the eigenvector-eigenvalue link} – the perfect correlation between the initial state $|{a_n}\rangle$ of the system $S$ and the final state $|{M_n}\rangle$ of the apparatus $M$.$\,$\footnote{\label{f5}Here it is also assumed the absence of degeneracy meaning there is only one eigenstate for each eigenvalue.\vspace{5pt}}\\

\noindent Let us return to the propositional formula (\ref{2}). Since the quantum evolution of the system $S+M$ is given as $|{\Psi}_t\rangle=\hat{U}(t) |{\psi}\rangle |{M_0}\rangle$, where $\hat{U}(t)$ stands for the time evolution operator, namely, $\hat{U}(t)=\exp(-itH_{S+M}/{\hbar})$, the predicate $P(|{\Psi}_t\rangle)$ is defined such\smallskip

\begin{equation} \label{4} 
     P(|{\Psi}_t\rangle)
     \equiv
     \bigg(
        \Big(
          |{\Psi}_t\rangle
           =
          \hat{U}(t) |{\Psi}_0\rangle
        \!\Big)
        \!\wedge\!
        \Big(\!
           |{\psi}\rangle = {c_1}|{a_1}\rangle + {c_2}|{a_2}\rangle
        \!\Big)
        \!\wedge\!
        \Big(
           \exists |{M_0}\rangle
           \!\in\!
           \mathcal{H} 
           \: P(|{M}_0\rangle)
        \!\Big)
     \bigg)
   \;\;\;\;   
\end{equation}
\smallskip

\noindent that it includes the statement $\left( |{\Psi}_t\rangle = \dots \right)$ asserting the validity of the Schr\"odinger dynamics for the measurement process, the statement $\left( |{\psi}\rangle = \dots \right)$ declaring the validity of the quantum superposition principle for the system $S$, and the statement $\left( \exists |{M_0}\rangle \dots \right)$ affirming the existence of the ready state of the apparatus $M$ as a quantum state vector $|{M}_0\rangle$, in which $\mathcal{H}$ denotes an abstract (separable, infinite-dimensional) Hilbert space and $P(|{M}_0\rangle)$ stands for the propositional function of the vector $|{M}_0\rangle$.\\

\noindent As follows, to assign a logical value to the proposition $\left( (\{a_1\}) \vee (\{a_2\}) \right)$ in advance of the measurement, the predicate $P(|{\Psi}_t\rangle)$ must be calculated. But to do so, a witness is required that can provide computational evidence supporting the statement $\left( \exists |{M_0}\rangle \dots \right)$ since this is the one in the formula (\ref{4}) that involves the existential quantification $\exists$.\\

\noindent Were the state vector $|{M}_0\rangle$ to exist, it would be a vector of the Hilbert space $\mathcal{H}_M \subseteq \mathcal{H}$ of the apparatus $M$ (as much as the state vectors $|{M}_1\rangle$ and $|{M}_2\rangle$ would be). From another side, the eigenvectors $|{M}_n\rangle$ of the Hamiltonian operator $H_M$ corresponding to the total energy of the apparatus $M$ must provide an orthonormal basis $\{ |{M}_n\rangle \}$ for the Hilbert space $\mathcal{H}_M$ they span. For this reason, let us consider \textit{the decision problem of the Schr\"odinger equation} $\Pi(H_M)$ for the Hamiltonian operator $H_M$\smallskip

\begin{equation} \label{5} 
     \Pi(H_M)
     \equiv
     \big(
        \{ |{M}_n\rangle \}
        \neq \emptyset 
     \big)
   \;\;\;\;   
\end{equation}
\smallskip

\noindent whose output is taken to be ‘true’ if the solution set $\{ |{M}_n\rangle \}$\smallskip

\begin{equation} \label{6} 
        \{ |{M}_n\rangle \}
        \equiv
        \left\{
        |{M_n}\rangle \!\in\! \mathcal{H} 
        \,\Big|\,
        i\hbar\frac{\partial}{\partial t}
        |{{M}_n}\rangle
        =
        H_M |{{M}_n}\rangle
        \right\}
   \;\;\;\;   
\end{equation}
\smallskip

\noindent (i.e., the set of all vectors $|{{M}_n}\rangle$ for which the Schr\"odinger equation with $H_M$ holds) is not empty and `false' otherwise.\\

\noindent Most generally, the decision problem of the Schr\"odinger equation can be presented in the form\smallskip

\begin{equation} \label{7} 
     \Pi(H_C)
     \equiv
     \big(
        \exists |{\mathit{u}}\rangle \in \mathcal{H} \: P( |{\mathit{u}}\rangle, H_C )
     \big)
   \;\;\;\;  ,
\end{equation}
\smallskip

\noindent where the predicate $P( |{\mathit{u}}\rangle, H_C )$ encloses the general time-dependent  Schr\"odinger equation\smallskip

\begin{equation} \label{8} 
     P( |{\mathit{u}}\rangle, H_C )
     \equiv
     \Big(\!
        i\hbar\frac{\partial}{\partial t}
        |{\mathit{u}}\rangle
        =
        H_C |{\mathit{u}}\rangle
     \!\Big)
   \;\;\;\;  ,
\end{equation}
\smallskip

\noindent in which $H_C$ symbolizes a completely arbitrary Hamiltonian operator (which means that the term $H_C$ is free for substitution for any element in the set of all allowable Hamiltonian operators $H$) and the vector $|{\mathit{u}}\rangle$ represents the exact solution to this equation.\\

\noindent Contrary to microsystems, a typical macroscopic object with its uncontrolled and unlimited degrees of freedom cannot be assigned a specific (i.e., explicit, detailed and unambiguous) Hamiltonian operator $H_M$. This means that the truthfulness of the statement $\left( \exists |{M_0}\rangle \!\in\! \mathcal{H} \: P(|{M}_0\rangle) \right)$ would be secured if and only if the solution set $\{|{\mathit{u}}\rangle \}$ of the Schr\"odinger equation was \textit{in no case} empty, that is, the decision problem of this equation had to be \textit{universally true} (i.e., had the ‘truth’ output for all physical systems), namely,\smallskip

\begin{equation} \label{9} 
     \Pi(H_C)
     \leftrightarrow
         \big(
            \exists |{M_0}\rangle \!\in\! \mathcal{H} \: P(|{M}_0\rangle)
         \big)
   \;\;\;\;   
\end{equation}
\smallskip

\noindent (where -- in line with the notion of constructiveness -- the symbol $\leftrightarrow$ may be viewed as an abbreviation of  ``can be replaced in a proof with''). \\

\noindent Let us consider the following claim\smallskip

\begin{equation} \label{10} 
     \Pi(H_C)
     = \top
   \;\;\;\;  .
\end{equation}
\smallskip

\noindent It is clear that this claim (logically equivalent to \textit{the assertion of the universal validity of the Schr\"odinger dynamics}) could be proved if the Schr\"odinger  equation was capable of being exactly solved by a generic algorithm, i.e., for the Hamiltonian operator $H_C$.\\

\noindent However, such an algorithm does not exist (at least as it is known in the present state of our knowledge \cite{Popelier}). To be exact, the decision problem $\Pi(H_C)$ is known to be undecidable, i.e., there does not exist a single (generic) method that can in a finite number of steps correctly solve this problem for any allowable Hamiltonian operator $H$ (even though methods of finding the solution set $\{|{\mathit{u}}\rangle \}$ of the Schrödinger equation with some particular $H$ are known).\\

\noindent Indeed, as it is argued in \cite{Bolotin}, the decision problem of the Schr\"odinger equation is parallel to \textit{the general spectral gap problem} that asks whether a given Hamiltonian operator $H$ has a spectral gap (i.e. the energy difference between the ground state and the first excited state of the system in the thermodynamic limit). To be sure, if there were a generic algorithm capable of obtaining the solution set $\{|{\mathit{u}}\rangle \}$ of the Schr\"odinger equation for all allowable Hamiltonian operators $H$, then such an algorithm would be able to answer not only the question whether this solution set $\{|{\mathit{u}}\rangle \}$ is empty or not but also the question whether the spectrum of the eigenvalues corresponding to $\{|{\mathit{u}}\rangle \}$ is discrete and gapped or continuous and gapless. Yet, according to the result of the paper \cite{Cubitt}, \textit{the general spectral gap problem is undecidable}. One can infer from this conclusion that the generic algorithm for solving the Schr\"odinger equation with an arbitrary $H_C$ does not exist and consequently the decision problem  $\Pi(H_C)$ is undecidable.\\

\noindent In order to ensure the general undecidability of this problem, in the intuitionistic interpretation of quantum mechanics the declaration of the universal validity of the Schr\"odinger dynamics is \textit{weakened} (in comparison with other interpretations that are based on classical or quantum logic). Specifically, it is asserted that unless one has a proof that for the particular system $\Pi(H) = \top$, the Schr\"odinger dynamics can be considered only being \textit{possibly valid} for this system. Symbolically, this can be written down as the following entailment\smallskip

\begin{equation} \label{11} 
     \Big(
        \exists H \: \Pi(H)
        = \top
     \Big)
     \vdash
     \Big(
        \Diamond  \Pi(H_C)
        = \top
     \Big)
   \;\;\;\;  ,
\end{equation}
\smallskip

\noindent where $\Diamond$ stands for the modal operator of possibility \cite{Chellas}. This entailment assigns the truth value $\top$ to the proposition ``it is possible that $\Pi(H_C)$'' as a logical consequence of the fact that for some Hamiltonian operators $H$ the solution set of the Schr\"odinger equation can be explicitly demonstrated$\,$\footnote{\label{f6}In contrast to the intuitionistic interpretation, the RS/MWI and the decoherence program infer the universal validity of the Schr\"odinger dynamics from the same fact, i.e., $(\exists H \: \Pi(H)=\top) \vdash (\Pi(H_C)=\top)$. Indeed, as it is stated for example in \cite{Leggett}, since ``there is satisfactory and often excellent evidence'' that the quantum mechanical (QM) framework ``is quantitatively valid'' and at the same time ``there is, at least at present, no positive experimental evidence that it is not valid in other regions where it has not been directly tested'', then ``the principle of Occam's razor would certainly suggest that the intellectually economical attitude is to assume that the general conceptual scheme embodied in QM is in fact valid for the whole of the physical universe without restriction''.\vspace{5pt}}.\\

\noindent By the axioms of seriality and reflexivity of modal logic, i.e., $\Box \Pi(H_C) \rightarrow \Diamond  \Pi(H_C)$ and $\Box \Pi(H_C) \rightarrow \Pi(H_C)$ respectively (where $\Box$ is the operator of necessity), the equality $\Diamond  \Pi(H_C) = \top$ suggests  $\Pi(H_C) = \top$ but it equally may be $\Pi(H_C) = \bot$. This means that the entailment (11) declares \textit{contingency} of the decision problem $\Pi(H_C)$ (in other words, the outcome of  $\Pi(H_C)$ is asserted to be uncertain). As a result, from the formula (\ref{9}) one infers the following evaluation\smallskip

\begin{equation} \label{12} 
      \big(
         \exists |{M_0}\rangle \!\in\! \mathcal{H} \: P(|{M}_0\rangle)
      \big)
      =
      \{\}
   \;\;\;\;  ,
\end{equation}
\smallskip

\noindent where symbol $\{\}$ represents the lack of the logical values. The meaning of the evaluation (\ref{12}) is that the statement asserting the existence of the state vector $|{M_0}\rangle$ cannot be decided computationally, i.e., by means of the Schr\"odinger equation.\\

\noindent This implies that the predicate $P(|{\Psi}_t\rangle)$ presented in the formula (\ref{4}) simply \textit{cannot be known}: Its truth or falsehood cannot be analyzed (computed). The unknowingness of the predicate $P(|{\Psi}_t\rangle)$ in turn implies that it is impossible to assign a definite logical value to $\left( (\{a_1\}) \vee (\{a_2\}) \right)$ -- and so to the proposition $(\{a_1,a_2\})$ -- in advance of the actual measurement of the observable $A$. In other words, because of the undecidability of the decision problem $\Pi(H_C)$, the logical values of the propositions of elementary events $(\{a_1\})$ and $(\{a_2\})$ cannot be decided (and thus exist) ahead of the measurement.\\

\noindent In passing, let us note that \textit{postulating the the universal validity} of the Schr\"odinger dynamics and in this way the existence of the ready state of the apparatus $M$ as a state vector  $|{M_0}\rangle$, one gets a contradiction known as \textit{the quantum measurement problem} (or, to be more exact, the problem of definite outcomes). Truly, when the both statements $\left( \exists |{M_0}\rangle \!\in\! \mathcal{H} \: P(|{M}_0\rangle) \right)$ and  $\left( \psi = {c_1}|{a_1}\rangle + {c_2}|{a_2}\rangle \right)$ are presumed to be true, the statement asserting the validity of the Schrödinger dynamics for the measurement process, $( |{\Psi}_t\rangle = \hat{U}(t) |{\Psi}_0\rangle )$, cannot be true: The linearity of the Schrödinger equation entails that in this case the initial state of the composite system $S+M$ evolves into the following final state\smallskip

\begin{equation} \label{13} 
     \big(
        {c_1}|{a_1}\rangle + {c_2}|{a_2}\rangle
     \big)
     |{M_0}\rangle
     \stackrel{t}{\longrightarrow} \:
     {c_1}|{a_1}\rangle |{M_1}\rangle
     +
    {c_2}|{a_2}\rangle |{M_2}\rangle
  \;\;\;\;    ,
\end{equation}
\smallskip

\noindent in which the symbol + representing the linear superposition cannot be replaced by the coordinating conjunction ``or'' such that $|{\Psi}_t\rangle = |{a_1}\rangle |{M_1}\rangle$ \textit{or} $|{\Psi}_t\rangle = |{a_2}\rangle |{M_2}\rangle$ (meaning the inclusive or exclusive disjunction). So, without supplying an additional postulate (for example, the wave-packet reduction postulate) or giving a suitable interpretation of the superposition ${c_1}|{a_1}\rangle |{M_1}\rangle + {c_2}|{a_2}\rangle |{M_2}\rangle$, it is impossible to explain the definite pointer position $|{M_n} \rangle$ corresponding to the value $a_n$, that is, the logical value `true' of the proposition $(\{a_1,a_2\})$, which is always perceived as the outcome of the actual measurement of the observable $A$.\\

\noindent Let us now evaluate the intuitionistic (Heyting) negation of the predicate $P(|{\Psi}_t\rangle)$:\smallskip

\begin{equation} \label{14} 
     \neg P(|{\Psi}_t\rangle)
     \equiv
     \neg\neg
     \bigg(
        \Big(\!
          |{\Psi}_t\rangle
          \neq
           \hat{U}(t) |{\Psi}_0\rangle
        \!\Big)
        \!\vee\!
        \Big(\!
              \psi \neq {c_1}|{a_1}\rangle + {c_2}|{a_2}\rangle
        \!\Big)
        \!\vee\!
        \neg
        \Big(\!
           \exists |{M_0}\rangle
           \!\in\!
           \mathcal{H} 
           \: P(|{M}_0\rangle)
        \!\Big)
     \bigg)
   \;\;\;\;  .
\end{equation}
\smallskip

\noindent From the definition of the decision problem $\Pi(H_C)$ it follows that its negation, namely, $\neg \Pi(H_C) \equiv \left( \forall |{\mathit{u}}\rangle \in \mathcal{H} \: \neg P( |{\mathit{u}}\rangle, H_C ) \right)$, is not constructively provable: In order to produce direct evidence of the truthfulness of this negation, one would have to demonstrate falsity of the Schr\"odinger equation, that is, $\neg P( |{\mathit{u}}\rangle, H_C ) \equiv \left( i\hbar\frac{\partial}{\partial t} |{\mathit{u}}\rangle \neq H_C |{\mathit{u}}\rangle \right) = \top$, \textit{for all} vectors $|{\mathit{u}}\rangle$ in the abstract infinite dimensional Hilbert space $\mathcal{H}$.\\

\noindent Instead, let us consider the double negation introduction\smallskip

\begin{equation} \label{15} 
   \neg \Pi(H_C)
   \leftrightarrow
   \neg
   \big(
           \neg \neg \Pi(H_C)
   \big)
   \;\;\;\;  ,
\end{equation}
\smallskip

\noindent which is a theorem in constructive logic. Since to prove $\neg \neg \Pi(H_C)$ intends to show that the solvability of the general Schr\"odinger equation would not be contradictory, the double negation $\neg \neg \Pi(H_C)$ has `true' output when there is no evidence against universal validity of Schr\"odinger dynamics. Otherwise stated, $\neg \neg \Pi(H_C) = \top$ holds if $\Pi(H_C)$ cannot be falsified or it is possibly true.\\

\noindent As it has been already noticed, the truthfulness of $\neg \neg \Pi(H_C)$ (i.e., $\Diamond \Pi(H_C) = \top$) can be inferred from the existence of any known exactly solvable quantum model. Thus, from (\ref{15}) it follows that\smallskip

\begin{equation} \label{16} 
   \neg \Pi(H_C)
   =
   \bot
   \;\;\;\;  .
\end{equation}
\smallskip

\noindent The last inference means that in the intuitionistic interpretation \textit{not-Schr\"odinger dynamics} is considered \textit{impossible} (i.e., false and necessarily false, that is, $\Box \neg \Pi(H_C) = \bot$).$\,$\footnote{\label{f7}It also generates $\Pi(H_C) \!\vee\! \neg \Pi(H_C) \neq \top$, which means that in the presented intuitionistic interpretation of quantum mechanics, the law of excluded middle is not admitted as an axiom.\vspace{5pt}}\\

\noindent Taking into account the negation of the equivalence (\ref{9}), this impossibility yields\smallskip

\begin{equation} \label{17} 
   \Big(
      \neg \Pi(H_C) = \bot
   \Big)
     \vdash
     \Big(
        \neg
        \big(
            \exists |{M_0}\rangle \!\in\! \mathcal{H} \: P(|{M}_0\rangle)
        \big)
        = \bot
      \Big)
   \;\;\;\;  .
\end{equation}
\smallskip

\noindent At the same time, in view of the deduction $\left( \neg \Pi(H_C) = \bot \right) \vdash \left( \neg P( |{\mathit{u}_C}\rangle, H_C ) = \bot \right)$, where $|{\mathit{u}_C}\rangle$ denotes a completely arbitrary vector in $\mathcal{H}$,  and the equivalence (holding true for small $t$ or if the Hamiltonian operator $H_{S+M}$ does not depend on $t$)\smallskip

\begin{equation} \label{18} 
   \Big(
          |{\Psi}_t\rangle
          =
           \hat{U}(t) |{\Psi}_0\rangle
   \Big)
   \leftrightarrow
   P( |{\Psi}\rangle, H_{S+M} ) 
   \;\;\;\;  ,
\end{equation}
\smallskip

\noindent where the propositional function $P( |{\Psi}\rangle, H_{S+M} )$ encloses the Schr\"odinger equation for $|{\Psi}\rangle$ and $H_{S+M}$, one infers that\smallskip

\begin{equation} \label{19} 
   \Big(
          \neg P( |{\mathit{u_C}}\rangle, H_C) 
          =
          \bot
   \Big)
   \vdash
   \Big(
      \big(
          |{\Psi}_t\rangle \neq \hat{U}(t) |{\Psi}_0\rangle
      \big)
      =
      \bot
   \Big)
   \;\;\;\;  .
\end{equation}
\smallskip

\noindent The expressions (\ref{17}) and (\ref{19}) together with the evaluation $\left( \psi \neq {c_1}|{a_1}\rangle + {c_2}|{a_2}\rangle \right) = \bot$ (which is trivial if the system $S$ is microscopic$\,$\footnote{\label{f8}This evaluation is due to the easily verifiable assumption that the microsystem $S$ can be prepared in two different eigenstates of the observable $A$ and in a superposition of two such states.\vspace{5pt}}), result in that the negation $\neg P(|{\Psi}_t\rangle)$ must be false for any predicate variable $|{\Psi}_t\rangle$.\\ 

\noindent As follows, even though the propositional formula (\ref{2}) itself does not have an a priori logical value,  i.e., $((\{a_1\}) \vee (\{a_2\}))=\{\}$, the negation of this formula must be `false': $\left( \neg \left(\{a_1\}\right) \wedge \neg \left(\{a_2\}\right) \right) = \bot$.\\

\noindent Furthermore, since the predicate $P(|{\Psi}_t\rangle)$ of the propositional formula for the second part of the compound expression (\ref{1}), namely, $\left( \neg \left(\{a_1\}\right) \vee \neg \left(\{a_2\}\right) \right)$, is the same as one presented in the formula (\ref{4}), that is,\smallskip

\begin{equation} \label{20} 
        \big(
           \neg \left(\{a_1\}\right) \vee \neg \left(\{a_2\}\right)
        \big)
        \equiv
        \left(
           |{\Psi}_t\rangle
           \neq
          |{a_1}\rangle  |{M_1}\rangle
          \vee
          |{\Psi}_t\rangle
           \neq
          |{a_2}\rangle  |{M_2}\rangle
        \,\Big|\,
          P(|{\Psi}_t\rangle)
        \right)
   \;\;\;\;  ,
\end{equation}
\smallskip

\noindent the negation of this part must be `false' too:  $\left( \neg\neg \left(\{a_1\}\right) \wedge \neg\neg \left(\{a_2\}\right) \right) = \bot$.\\

\noindent So, if the propositions $(\{a_n\})$ can be assigned logical values and \textit{for that reason} double negation elimination $\neg\neg (\{a_n\}) \leftrightarrow  (\{a_n\})$ can be applicable to them, one will get the following equalities: $\left( \left(\{a_1\}\right) \vee \left(\{a_2\}\right) \right) = \top$ and $\left( \left(\{a_1\}\right) \wedge \left(\{a_2\}\right) \right) = \bot$.\\

\noindent Together these two equalities mean that after the measurement of the observable $A$ -- that is, in the moment when the records of the logical values of $(\{a_1\})$ and $(\{a_2\})$ are created -- these propositions turn out to be \textit{mutually exclusive}\smallskip

\begin{equation} \label{21} 
        (\{a_1,a_2\})
        \equiv
        \big(
           \left(\{a_1\}\right) \oplus \left(\{a_2\}\right)
        \big)
        = \top
   \;\;\;\;  .
\end{equation}
\smallskip

\noindent Apropos of the normalization requirement ${c_1}^2+{c_2}^2=1$, let us observe that it would match the operation of exclusive disjunction (\ref{21}), if the squared norms ${c_1}^2$ and ${c_2}^2$ were viewed as the elements of the Boolean domain $\{\top,\bot\} \equiv \{1,0\}$ corresponding respectively to the logical values of the propositions $(\{a_1\})$ and $(\{a_2\})$ once these values were created in the measurement of $A$.\\

\noindent As regards the squared norms involved, according to Gleason's theorem (modified for the case of two-dimensional state spaces \cite{Zela}), if one would like to assign a real valued function $m(\{a_n\}) \geq 0$ -- e.g., a post-measurement logical value of the proposition $(\{a_n\})$, namely, $m(\{a_n\})=(\{a_n\}) \in \{1,0\}$ -- to the vector $|{a_n}\rangle$ of the orthonormal set $\{ |{a_1}\rangle, |{a_2}\rangle \}$ such that the following sum\smallskip

\begin{equation} \label{22} 
        m(\{a_1,a_2\})
        =
        m\big(
               \left(\{a_1\}\right) \vee \left(\{a_2\}\right)
            \big)
        =
        m\left(\{a_1\}\right) + m\left(\{a_2\}\right)
        =
        1
   \;\;\;\;   
\end{equation}
\smallskip

\noindent holds true whenever the equality entailing the mutual exclusiveness of the outcome events $\{a_1\}$ and $\{a_2\}$\smallskip

\begin{equation} \label{23} 
        m\big(
               \left(\{a_1\}\right) \wedge \left(\{a_2\}\right)
            \big)
        =
        m\left(\{a_1\}\right) \times m\left(\{a_2\}\right)
        =
        0
   \;\;\;\;   
\end{equation}
\smallskip

\noindent is valid, then the only possible choice is $m(\{a_n\}) = {| \langle {a_n}|{\psi}\rangle |}^2$ in which $|{\psi}\rangle$ is an arbitrary but fixed vector $|{\psi}\rangle = c_1 |{a_1}\rangle + c_2 |{a_2}\rangle$.\\

\noindent For the sake of clarity and simplicity, let us suppose that $|{\psi}\rangle$ is the state of \textit{an equal superposition} (i.e., a superposition in which all the coefficients have equal norms) such that this state can be written down as $|{\psi}\rangle = 1/\sqrt{2} ( e^{i{\phi}_1}|{a_1}\rangle + e^{i{\phi}_2}|{a_2}\rangle )$ where phases ${\phi}_1$ and ${\phi}_2$ are real. As it can be readily seen, in that case the eigenstates $|{a_1}\rangle$ and $|{a_2}\rangle$ are interchangeable in  $|{\psi}\rangle$ up to the phase factors of the superposition coefficients $c_1$ and $c_2$, that is, $e^{i{\phi}_1}|{a_1}\rangle \leftrightarrows e^{i{\phi}_2}|{a_2}\rangle$.$\,$\footnote{\label{f9}Such an interchangeability results by letting the unitary ``swapping'' operator $\hat{S}=e^{i{\phi_1}}|{a_1}\rangle\langle{a_2}|e^{-i{\phi_2}}+e^{i{\phi_2}}|{a_2}\rangle\langle{a_1}|e^{-i{\phi_1}}$ act on $|{\psi}\rangle$. This operator satisfies the invariance relation $\hat{S}|{\psi}\rangle=|{\psi}\rangle$.\vspace{5pt}}\\

\noindent On the other hand, the exclusive disjunction ${|c_1|}^2+{|c_2|}^2=1$ demonstrates that only the magnitudes of the superposition coefficients $c_1$ and $c_2$ can correspond to the logical values $\{1,0\}$ of the propositions of elementary events $(\{a_1\})$ and $(\{a_2\})$ once these values are created. That is, the phase factors $e^{i{\phi}_1}$ and $e^{i{\phi}_2}$ do not indicate which one of these propositions become 1. Accordingly, the symmetry  $e^{i{\phi}_1}|{a_1}\rangle \leftrightarrows e^{i{\phi}_2}|{a_2}\rangle$ possessed by the equal superposition state $|{\psi}\rangle$ implies that prior to the measurement the propositions $(\{a_1\})$ and $(\{a_2\})$ are \textit{indistinguishable from one another} except for their names, i.e., $(\{a_1\}) \leftrightarrows (\{a_2\})$.\\

\noindent So, considering that the equal superposition state $|{\psi}\rangle$ contains no physical evidence favoring one proposition over another, one can invoke \textit{the principle of indifference} \cite{Hajek12} and deduce from the a priori nonexistence of the logical values of $(\{a_1\})$ and $(\{a_2\})$ in the state $|{\psi}\rangle$ that both of these propositions are \textit{equally likely} to become 1 during the measurement.$\,$\footnote{\label{f10}This inference is in line with Jayne's’ invariance condition \cite{Jaynes} according to which \textit{equal probabilities should be assigned to equivalent propositions}.\vspace{5pt}}\\

\noindent In a word, due to the fact that the propositions of elementary events $(\{a_1\})$ and $(\{a_2\})$ are indistinguishable (equivalent) in the equal superposition state $|{\psi}\rangle$ and after the measurement correspond to the mutually exclusive events such that $(\{a_1\}) + (\{a_2\}) = 1$, one can assign both of these propositions \textit{an equal probability} $\mathrm{Pr}(\{a_1\}) = \mathrm{Pr}(\{a_2\}) = 1/2$ of coming out 1 in the course of the measurement, inasmuch as $(\{a_1\})$ and $(\{a_2\})$ cannot be evaluated before the measurement.$\,$\footnote{\label{f11}The premeasurement logical values of the propositions $(\{a_n\})$ cannot be found (computed) by means of the Schr\"odinger equation and since \textit{not-} (i.e., \textit{other-than-}) Schr\"odinger dynamics is impossible, the probabilities of the events $\{a_n\}$ are interpreted as corresponding to \textit{a genuine stochastic process}.\vspace{5pt}}\\

\noindent This concludes the presentation of the way in which irreducible probabilities – a central element of the Born rule – appear in the intuitionistic interpretation of quantum mechanics in the case of the superposition coefficients of equal magnitude and the two-dimensional state space of the measured system $S$.\\

\noindent The generalization of the presented account to the case of $N$ coefficients with equal norms is rather straightforward$\,$\footnote{\label{f12}Strictly speaking, the operation of exclusive disjunction, XOR, is true when an odd number of propositions $(\{a_n\})$ are true. So, in the case, in which the dimension $N$ of the sample space of the measurement $\Omega$ is greater than 2, the proposition of a definite outcome must contain the additional to the operation XOR term $B_N$ providing for the true output when only a single $(\{a_n\})$ is true. For example, when $N=4$, this proposition takes the form $({\Omega})=(\mathrm{XOR}_4 \wedge \neg B_4)$, where $\mathrm{XOR}_4 = ((\{a_1\}) \oplus (\{a_2\}) \oplus (\{a_3\}) \oplus (\{a_4\}))$ and $B_4 = \mathrm{max} ({\{(\{a_i\}) \times (\{a_j\}) \times (\{a_k\})\}}_{i>j>k \in \{1,2,3,4\}})$.\vspace{5pt}}. As to the extension of the account to the case of the superposition $|{\psi}\rangle=\sum_{n=1}^{N} {c_n}|{a_n}\rangle$ with the coefficients $c_n$ of non-equal magnitudes, it can be achieved by means of logical theories of probability (see, for example, \cite{Carnap, Vohanka}) which (in order to save the principle of indifference and extend it to the case of unsymmetrical evidence) assert that the elementary events $\{a_n\}$ may be assigned unequal \textit{weights} $w_n$ (that are nothing more than replication counts indicating duplicated observations) and probabilities can be computed whatever the evidence may be, symmetrical or not.\\

\noindent Along these lines, let us suppose that the squared norms of the coefficients $c_n$ (of non-equal in general magnitudes) can be expressed as a fraction of two positive integers, namely, $|{c_n}|^2={w_n}/(\sum_{n=1}^{N} {w_n})$. Then, the proposition of elementary event $(\{a_n\})$ can be brought into the following formal form\smallskip

\begin{equation} \label{24} 
        (\{a_n\})
        \equiv
        \big(
               \left({\{a_n\}}_1\right) \vee \left({\{a_n\}}_2\right) \dots \vee \left({\{a_n\}}_{w_n}\right)
        \big)
        =
        \sum_{i=1}^{w_n} \left({\{a_n\}}_{i}\right)
   \;\;\;\;    ,
\end{equation}
\smallskip

\noindent where the subscript $i$ indicates replicated disjoint events ${\{a_n\}}_{i}$ which are all identical with respect to $(\{a_n\})$. The association of the propositions $(\{a_n\})$ with the a priori probabilities $\mathrm{Pr}(\{a_n\})=1/N$ has been already established in the special case where all the possible events $\{a_n\}$ have unity weights $w_n$. Thus, in the case of the weight $w_n>1$, a similar line of reasoning can be used to prove the next formula\smallskip

\begin{equation} \label{25} 
        \mathrm{Pr}(\{a_n\})
        =
        \sum_{i=1}^{w_n} \mathrm{Pr}\big({\{a_n\}}_{i}\big)
        =
        \frac{w_n}{\sum_{n=1}^{N} {w_n}}
   \;\;\;\;    .
\end{equation}
\smallskip

\noindent As follows, the presented account is restricted to a finite dimensional sample space $\Omega$ and thus a finite dimensional state space. But even if a Hilbert space is countably infinite, the given account may still be held up whenever a replacement of a countably infinite orthonormal basis of the measured system $S$ with some truncated (finite-dimensional) basis can be justified. \textit{Such is the case of a typical microsystem} for which one can safely assume a discrete spectrum of energies limited by some finite upper level whose order of magnitude is similar to ones of energies of electrons in an atom or a solid.\\

\noindent However, the shown intuitionistic account may not be extended to an uncountable (i.e., uncountably infinite) Hilbert space \textit{which is the case of a typical macroscopic system} whose degrees of freedom – in particular the macroscopic ones – can vary continuously and in an unconfined, unbounded manner. Apparently, this is equivalent to the elimination of the applicability of quantum probabilities and the Born rule to common macroscopic systems.\\

\noindent But then again, due to the equivalence of boundedness and continuity \cite{Hellman}, unbounded linear Hermitian operators in an uncountable Hilbert space are discontinuous and thus generate \textit{noncomputability} \cite{Myrvold95}. This means that the application of quantum formalism to a typical macroscopic system necessarily causes the existence of mathematical entities in the theoretical representation of the system that are incapable of being computed by any deterministic algorithm in any finite amount of time.$\,$\footnote{\label{f13}To be precise, one can no longer speak of the Hilbert space of a typical macroscopic system since ``the observables of an infinite system usually have a host of physically inequivalent representations, corresponding to macroscopically different classes of states'' \cite{Mundici}.\vspace{5pt}} Clearly, those entities cannot be verifiable (demonstrable) which makes the quantum mechanical description of any typical macroscopic system constructively unprovable and for that reason inapplicable.\\

\section{Conclusion}\label{Conclusion}

\noindent The essential question that may raise now is whether probabilities appearing in the intuitionistic interpretation are objective (i.e., fundamentally ingrained in nature) or subjective (i.e., due to the rational agent's ignorance). The peculiarity here is that neither answer to this question can be regarded as a correct one.\\

\noindent Really, according to the explanation presented in the previous section, quantum probabilities are introduced to account for the agent's lack of knowledge about the premeasurement logical values of the propositions of elementary events $(\{a_n\})$ in the case where the state $|{\psi}\rangle$ does not belong already (i.e., before the act of measurement) to one of the vectors $|{a_n}\rangle$ of the eigenbasis of the observable $A$ that is going to be measured (and so the sample space of the measurement $\Omega$ contains more than one element). As long as these propositions are considered symmetrical with respect to the state $|{\psi}\rangle$, the agent has no other information than the number of the mutually exclusive events $\{a_n\}$ that can occur during the measurement. Consequently, the agent is justified in assigning each of the possible events the equal probability. In this manner, the resultant probabilities may be called \textit{subjective} seeing as they reflect the agent's incomplete knowledge of the world.\\

\noindent But, on the other hand, quantum probabilities are \textit{objective} in that they are a consequence of the fact that the decision problem of the Schrödinger equation $\Pi(H_C)$ is undecidable. That is, for this problem it is impossible to construct an algorithm (applicable to all systems, i.e., all allowable Hamiltonian operators $H$) that would always lead to a correct `true'-or-`false' answer. Hence, the origin of the undecidability of the problem $\Pi(H_C)$ is not subjective, i.e., it is not just a lack of agent's imagination: No matter how ingenious the rational agent is, there is no way to exactly solve the Schr\"odinger equation for any given physical system. As a result, the validity of the linear, deterministic Schr\"odinger evolution for an arbitrary system – such as a typical macroscopic apparatus $M$ – cannot be proven constructively, i.e., by demonstrating a procedure of finding the solution set  $\{|{\mathit{u}}\rangle \}$ of the Schr\"odinger equation for a completely arbitrary Hamiltonian operator.\\

\noindent The similar ambiguity characterizes the question whether or not the formalism of quantum mechanics – in agreement with the intuitionistic interpretation – can be considered complete. Definitely, as stated by the intuitionistic interpretation, the quantum description of reality can be considered \textit{incomplete} as it contains the proposition $(\{a_1,a_2\})=\{\}$ which is undecidable from within theory (the truth value of this proposition cannot be predicted through calculations with the Schr\"odinger equation), but which is, nonetheless, decidable through experiment (i.e., the measurement of the observable $A$). And yet, no further accumulation of information about the composite system $S+M$ could help the agent to decide (compute) the logical values of the propositions $(\{a_1\})$ and $(\{a_2\})$ before the measurement, inasmuch as the fact of the statement $( \{|{\mathit{u}}\rangle \} \neq \emptyset )$ being neither provable nor disprovable is not determined by any finite amount of the information available to the agent. In view of that fact, the presented intuitionistic account of quantum description may be regarded as \textit{complete}.\\

\bibliographystyle{References}
\bibliography{References}

\end{document}